# A survey on flexible/restricted skyline and their applicability

Davide Canali

Politecnico di Milano
Milan, Italy
davide1.canali@mail.polimi.it

**Abstract**

Skyline and Top-k are two of the most important methods to extract information from datasets, but both come with their drawbacks, that's why lately some new techniques that try to mix the features of the two have been studied. In this survey three new operators are analysed, $\mathcal{F}$-Skyline, ORU/ORD, and $\varepsilon$-Skyline. After giving the main ideas behind those and their properties, they are compered on 3 fundamental features such as personalization, cardinality control, and generalization to guide the user to choose the best one for any task.

## 1 Introduction

We are living in a world full of data and every day an overwhelming amount of those are acquired and stored to retrieve useful information, but with that quantity of data doing so, it's not a trivial task. That's why since data have become more and more vital to the business of a company, techniques for finding the right information in a database have been studied more deeply.

Information is not the same for everyone, different clients may have different interests so it's not possible to automatically compute the best for all of them, but on the other hand, having a general view of the most interesting data in a database could be very useful. That's why two main techniques emerged in this field: top-k query which by using a scoring function can find the top k tuple based on their scoring. This technique can well manage the problem of client preference but it's not possible to compute those without knowing the said function from the user. Skyline instead is the other approach which it's based on the notion of dominance, a detailed description of dominance will be provided later, and the skyline result is all the tuples that are not dominated by any other tuple. It's easy to see that since no external input is required, this operation can be done automatically, the main drawback of skylines is the fact that the cardinality of the result cannot be controlled and that means that possibly the result of the skyline could still be unhelpful for the client since the quantity of information is still unmanageable.

To overcome those problems different techniques have been proposed and flexible Skyline [1] is one of those, the idea behind $\mathcal{F}$-Skyline is quite simple, by using some form of constraints over the domain space this technique can consider different attributes with different priorities and with that reflect some form of user preference as in top-k, that improves also the cardinality of the result since tighter constraints mean a more restricted result set. ORD and ORU [2] are the other two approaches to this problem which both resolve completely the cardinality problem by adding a parameter that specifies the number of expected tuples in the final list and, by a user-specific function can order the importance of different attributes. Last ε-Skyline [3] will be described, this technique uses a set of weights on the attributes and a constant ε to find the non-dominated tuples and reduce the result dimension.

# 2 Top-K Query

In this section a brief description of the most important properties of top-k queries will be given, mainly to highlight the aspects that will be useful to describe flexible/restricted skylines.

As said the main idea behind top-k is to use a scoring function to order tuples in a list and then retrieve the first k elements of that list, both the scoring function and the parameter k are given by the user and so they respect its preferences.

## 2.1 Scoring function

A scoring function maps some attributes of the object to a single value that represents its worth, which is how much that object meets the user preferences, this is the key component of this technique since it will be used to order the tuples.

The properties of the function are an important factor in implementing the top-k technique and so there are different types of scoring functions [4]:
- Monotone Function
- Generic Function
- No Ranking Function

**Monotone Function**

It's the most used type of function since it not only models an important part of the scenarios in which top-k is useful [5], but also it adds important features that can be exploited to speed up the process. It is possible to compute an upper bound that all objects, not already seen, cannot pass, since we just need to use the last seen value of every attribute to compute a bound using the chosen monotone function. It's possible to have an upper bound for each object that has some attributes unknown by just using the last seen value for each unknown attribute, those properties allow to stop the process of searching the top-k tuple earlier. [6] [4]

**Generic Function**

Using non-monotone function brings some challenges since it's not possible to bound some tuple based on the order of appearance in the table, although some studies [7] [8] found some technique that made possible to use this kind of function. The main idea is to transform the top-k problem into an optimization problem based on the shortest pathfinding on a graph, using algorithms such as A*, or using R-B trees to index possible query answers and search in that space the best one.

**No Ranking Function**

In this case, the problem overlaps with the skyline problem, and it will be described in the next section since it's based on the concept of dominance and the goal is to get the more interesting tuples.

# 3 Skyline

In this section is described the concept of skyline and its properties, mainly the ones used in flexible/restricted skylines.

As already said the fundamental concept of skylines is dominance, that's used to retrieve the most interesting tuples in a dataset which are those whose is not dominated by anyone.

A formal definition of dominance is the following:

*Let A be the set of relevant attributes a tuple has $A = [a_1, a_2, \ldots, a_d]$, r be the set of tuples that are considered and let $t, s$ be two tuples $\in r$:*
$$t \prec s, iff\ \forall i, 1 \leq i \leq d \rightarrow t[A_i] \leq s[A_i], and\ \exists j, 1 \leq j \leq d\ \wedge t[A_j] < s[A_j]$$
*And so, the formal definition of Skyline over r is:*

$$Sky(r) = \{t \in r \mid \nexists s \in r, s \prec t\}$$

An important property of skyline is the fact that for every monotonic scoring function the tuple that scores the highest value is in the skyline, formally defined as follow:

$$\forall r, \forall t \in r, t \in Sky(r) \Leftrightarrow \exists f \in M. \forall t' \in r, t' \neq t \Rightarrow f(t) > f(t')$$
*Where M is the set of all monotone functions over the set of tuples.*

proofs of that are given by [9].
That's the main theorem that described the real utility of skyline which is to show the general best tuples of a database.

## 3.1 Classes of Skyline

Given the notion of scoring function the first overlap with ranking query can be obtained by other classes of skyline [10]:
- Constrained Skyline: the skyline operator is applied on a subset of tuples given by the result of a ranking.
- Ranked Skyline: the result of the skyline is ordered by some scoring function.
- Enumerating Skyline: the result of the skyline is ordered by a scoring function that returns the number of dominated tuples.
- K-dominating Skyline: an enumerating skyline but only the top-k tuples are returned as result.
- K-Skyband: return only the tuples that are dominated by at most k other objects.

## 3.2 Cardinality

As said one of the most challenging problems of using skyline is predicting the cardinality of the result, if it's too large this technique loses its utility, so this topic has been addressed in different studies [11] [9], starting with two main domain assumptions: (i) attributes are statistically independent between each other and (ii) the probability of two tuples to have the same value over an attribute is minimal.
Given those two assumptions we can estimate the cardinality of the skyline by the following recursive function:

$$l_{d,n} = l_{d,n-1} + \frac{1}{n} \cdot l_{d-1,n}$$
*Where $l$ is the cardinality, $d$ is the number of attributes and $n$ is the number of tuples.*

If we don't consider the second domain assumption, tuples may have attributes with the same value, and duplicated tuples can occur. The first effect is that the distribution over an attribute of the values is no more negligible, and in fact, the number of tuples in the skyline is reduced compared to the previous condition. The second effect that can be seen is that, if we partition values over an attribute, that is values in the same partition is consider as equal when testing dominance, this can produce two effects:
A tuple that before the partition dominated another tuple may not dominate it after the partitioning is applied, or the other way around, a tuple that didn't dominate a tuple, after the operation, it may dominate it. The larger the number of attributes is, the more the second case will occur and so limiting the number of dimensions result in a smaller set of tuples after the skyline operator.

# 4 Flexible Skyline

Flexible Skyline [1] tries to resolve the problem of inserting user preferences into the Skyline operator, to do so the concept of $\mathcal{F}$-dominance has been introduced, and with that two new skyline subsets can be derived, ND, which is the set of all non-$\mathcal{F}$-dominated tuples, and PO, which is the set of all the tuples that are potentially optimal, that means tuples that are the best for some function in $\mathcal{F}$. The two main benefits of this approach are the fact that introducing constraint on attributes means a more user-specific result, and the cardinality of the result set is reduced.

## 4.1 $\mathcal{F}$-Dominance

As said $\mathcal{F}$-dominance is the main concept behind this new skyline technique, and so this is a formal definition:

*Let $\mathcal{F}$ be a set of monotone scoring functions. A tuple t $\mathcal{F}$- dominates another tuples s ≠ t, denoted by*
$$t \prec_{\mathcal{F}} s, if \ \forall f \in \mathcal{F}. f(s) \geq f(t)$$

Since the definition of dominance is changed, the result of the skyline will be different too, and with that two new sets were created.

## 4.2 ND and PO sets

Given the definition of $\mathcal{F}$-dominance, we can now introduce the two sets of tuples given as a result of the operation, the non-dominated set, and the potentially optimal set.

In the ND set, there will be all the tuples that based on the notion of $\mathcal{F}$-dominance are not dominated by any other tuple:

$$nd(r; \mathcal{F}) = \{t \in r | \nexists s \in r. s \prec_{\mathcal{F}} t\}$$

Instead in the PO set, there will be all the tuples that for some scoring function in $\mathcal{F}$ have the highest score among all the tuples.

$$po(r; \mathcal{F}) = \{t \in r | \exists f \in \mathcal{F}. \forall s \in r. s \neq t \Rightarrow f(s) > f(t)\}$$

Using those sets can be given a user a good response to his request and criteria, much better than what's possible with just the normal Skyline operator.

To link this set with the standard skyline result we can enunciate a couple of properties:
1) Let $\mathcal{F}$ be a set of monotone scoring functions: $po(r; \mathcal{F}) \subseteq nd(r; \mathcal{F}) \subseteq Sky(r)$
2) Given two sets $\mathcal{F}_1$ and $\mathcal{F}_2$ such that $\mathcal{F}_1 \subseteq \mathcal{F}_2$: $po(r; \mathcal{F}_1) \subseteq po(r; \mathcal{F}_2)$ and $nd(r; \mathcal{F}_1) \subseteq nd(r; \mathcal{F}_2)$
3) Let MF be the set of all possible monotone scoring functions $po(r; \mathcal{F}) = nd(r; \mathcal{F}) = Sky(r)$

# 5 ORD and ORU

ORD and ORU [2] are two operators which aim to have a high personalization of the result set based on user's preferences and have a specified output size.

The idea behind those two new operators is that by using a vector w of weights, which correspond to the best-estimated user preferences, the best tuples are found by expanding this constraint in all directions and saving the found tuples until the desired output size has been reached.

## 5.1 Definitions

Still using r as the set of all tuples and A be the set of relevant attributes a tuple has $A = [a_1, a_2, ..., a_d]$, v is the weight (w) vector used to get the scoring of a tuple t:

$$S_v(t) = \sum_{i=1}^{d} a_i \cdot w_i$$

Let $\rho$ be the maximum distance between the weight vectors v and the estimated user preference vector w, such that for all v, $|v - w| \leq \rho$ and so, given two tuples t and s, if t scores (based on $S_v(\cdot)$) at least as high as s, for every vector v within distance $\rho$, and strictly higher for at least one of them, t $\rho$-dominates s, the tuples that are $\rho$-dominated by fewer than k others form the $\rho$-Skyband.

## 5.2 ORD

Given the seed vector w and the required output size m, ORD reports the records that are $\rho$-dominated by fewer than k others, for the minimum $\rho$ that produces exactly m records in the output.

### 5.2.1 Computing ORD

To find the result of the ORD we use the idea of inflection radius: if we take a tuple $t$ its inflection radius is that value of $\rho$ where t is $\rho$-dominated by less than k other tuples, to get that value we compute the k-Skyband in which, some tuple, the ones that dominate $t$ in the standard definition, will dominate $t$ for any $\rho$, and other, who do not dominate $t$ but score higher for w, for those tuples we can compute the radius $\rho$ for which they dominate $t$. Now we can see the value of the radius past which $t$ is dominated by fewer than k tuples.

With that, if we compute the entire k-Skyband, and for each tuple, we derive its inflection radius, we can then output the m tuples with the smallest radius.

Other approaches that don't require the computation of the entire k-Skyband are given in [2].

Given 3 tuples A, B, C each with 2 attributes Value and Rarity and a vector $w = [2, 1]$:

Example 1.1

|   | Value | Rarity |
|---|-------|--------|
| A | 10    | 10     |
| B | 15    | 4      |
| C | 20    | 0      |

We can see that the scores $S_w(t)$ are A=30, B=34, and C=40 so both B and C score higher than A for the seed vector but none of them dominates A in the classical definition of dominance.

We can compute the inflection radius of A given a $k = 2$ as follow:
Compute all the vectors $v_i$ of distance 1 from $w$: [1,1] [2,0] [3,1] [2,2].
By computing the score, using each vector, we can see that B doesn't dominate A, since its score using the $v$ vector [1,1] is 19, which is lower than 20, the score of A, this is sufficient to say that tuple A is dominated by less than 2 tuples for a radius of 1 which is its inflection radius.

## 5.3 ORU

Given the seed vector w and the required output size $m$, ORU reports the records that belong to the top-$k$ result for at least one preference vector within distance $\rho$ from w, for the minimum $\rho$ that produces exactly $m$ records in the output.

### 5.3.1 Computing ORU

The computation of the ORU is based on the convex hull, which is the smallest polytope that encloses all tuples in the dataset, with that, we call the upper hull or layer, the set of all facets of that convex hull whose normal vector is directed to the positive part of the plane, and last, we call top-region of t, the part of the domain in which t is the tuple that has the highest score. With that, we can compute ORU by recursively searching the next top-i result applying the property that, given a tuple t that is the top-i result, the next top-i + 1 for some preference vector v will be among the adjacent tuples in the same layer of t or in the tuples in the next layer whose top-region overlap the one of t.

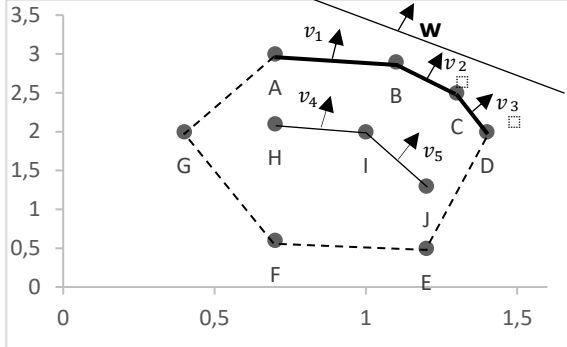
Example 2.1

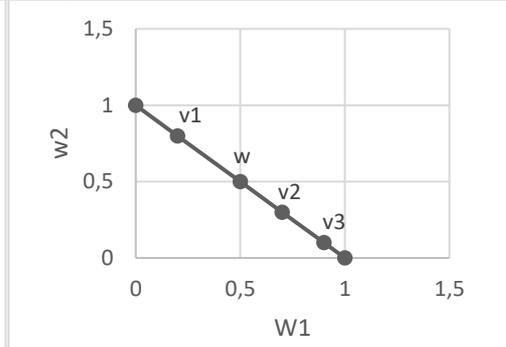
Example 2.2

In the example, the tuples of a dataset are shown on a graph and the upper hull is highlighted in bold, the facets of that hull are the segments A-B, B-C, C-D and all the corresponding normal vectors are shown.

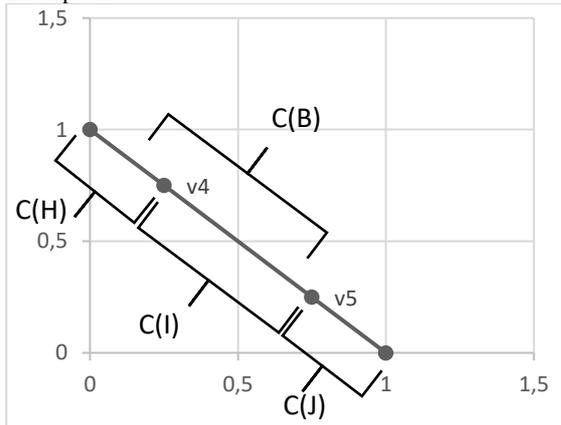
Example 2.3

Taken B as the top-i record, the record that scored the most given vector w, the next record that will score the maximum for a given vector v in $\rho$ distance from w, will be among the set of adjacent record {A, C} and the set of {H, I} which are the tuples that top-regions (C(H), C(I)) overlap the top-region C(B) which is the region between the normal vectors v1,v2.

# 6 $\mathcal{E}$-Skyline

ε-Skyline [3] is the third operator analysed in this survey, and like the other two techniques its goal is to incorporate user preferences into skyline, and, at the same time, add a cardinality control to the output list of tuples.

To do so, ε-Skyline uses both an array of weight, that represent the importance of each attribute, given by the user, and a constant ε which acts like a transposition of a tuple and so it can dominate more tuples (or less if ε<0).

## 6.1 ε-dominance

Given the basic idea of ε-Skyline we can extend the definition of dominance to incorporate both weights and the ε constant:

$$\text{Given a set of tuples with } d \text{ attributes and a set of weight } W = \{w_i \mid i \in [1, d], 0 < w_i \leq 1\},$$
$$\text{and a constant } \varepsilon \in [-1, 1], \text{for any two tuples } t_1 \text{ and } t_2$$
$$\text{is said that } t_1 \varepsilon\text{-dominate } t_2 \text{ denoted as } t_1 \prec_\varepsilon t_2$$
$$\text{if } \forall i \in [1, d], t_1[i] \cdot w_i \leq t_2[i] \cdot w_i + \varepsilon, \text{and } \exists j \in [1, d], t_1[j] < t_2[j]$$

Given this definition is easy to see the utility of the introduced constant $\varepsilon$, the fact that the value of each attribute of the second tuple is added up with $\varepsilon$ makes that, if two tuples are relatively similar to each other, only the best of the two is kept in the result, the similarity can be controlled by the weights and the $\varepsilon$ parameters, both given by the user.

## 6.2 Computing 𝓔-Skyline

The approach used to compute the ε-Skyline is based on a spatial index partition, separating tuples in regions, for each tuple t, we can compute its affecting region, in which every tuple will dominate the t for the given $\varepsilon$, with that, we can search for such tuple only in the region that intersects with the affecting region of t, if an entire edge of a region is inside the effecting region we know for sure that tuple t is dominated by at least another tuple and so it can be discarded, we can recursively apply this algorithm to find all the tuples in the ε-Skyline.

Example 3.1

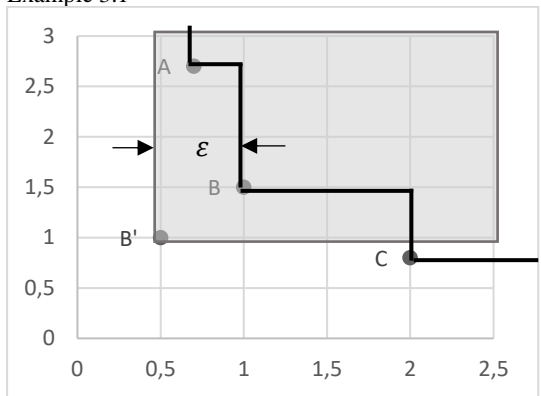

Given the dataset, with 4 tuples the graph shows how the $\varepsilon$ parameter changes the dominance relation between the tuples, B' now dominates A.

Example 3.2

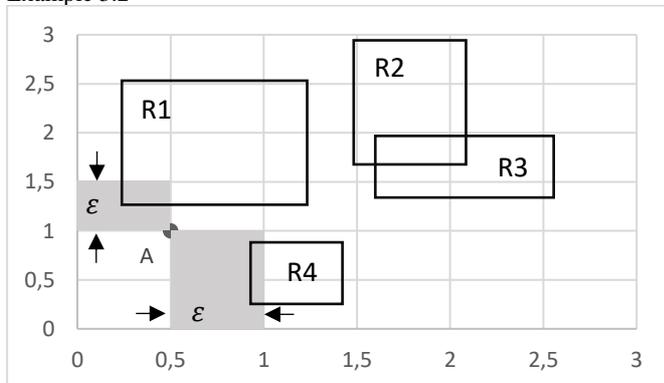

In this graph, each section R1-4 is the minimum region that contains a set of tuples and A is the chosen tuple to analyse, the grey areas are the regions in which if there is a tuple it will dominate the tuple A.

It's easy to see that regions 2 and 3 are dominated by A since their lower-left corner is dominated by A instead regions 1 and 4 need a further analysis since it's possible that a tuple in those regions is in the effecting region and so dominating A, in fact, in R4 there will be for sure a tuple that dominates A since one of its sides is inside the affecting region and by definition a tuple will be there.

# 7 Comparison

We compare the general properties of the 5 showed techniques, the 3 properties that will be analysed will be:
- Personalization: if it's possible to insert user preferences into the operator.
- Cardinality control: if there's some sort of control on the output size.
- Generalization: if the operator gives an overview of the entire dataset.

Table 1 – Features comparison

|  | Personalization | Cardinality Control | Generalization |
|---|---|---|---|
| Top-k | ✓ | ✓ | - |
| Skyline | - | - | ✓ |
| Flexible Skyline | ✓ | ✓ | ✓ |
| ORD/ORU | ✓ | ✓ | ✓ |
| ε-Skyline | ✓ | ✓ | ✓ |

All 3 techniques shown can merge the features of top-k and skyline, but to understand the real connection between those a deeper analysis is necessary.

## 7.1 Personalization

Regarding personalization all 3 can introduce the user preferences, but in different ways, $\mathcal{F}$-skyline uses multiple scoring functions to achieve that, instead, ORD/ORU use a set of weight on attributes, which can be translated as a scoring function itself, and ε-skyline does more than that and introduce even a constant. We can see that $\mathcal{F}$-skyline can focus more specifically on a section of the domain, more than is possible with the other two, since combining multiple constraints allow to partition the space more freely.

The only constraint about the scoring function used by all those techniques is that it must be monotone.

To obtain the weight and so the scoring function it's possible to use some mining techniques [2] and this is possible if we are talking about ORD/ORU and ε-skyline but with $\mathcal{F}$-skyline this is not that simple since $\mathcal{F}$ can contain an infinite number of those functions.

All these operators trade the feature of personalizing the skyline with having higher complexity than top-k query, which remains the preferred choice if a fast and user-specific analysis of the database is the goal, or a non-monotone scoring function is required.

## 7.2 Cardinality Control

If we analyse the control over the output size, ORD/ORU clearly have an advantage, since the cardinality is chosen via a parameter, it's fixed, just like top-k query.

The other two reduce the output dimension indirectly, thanks to constraints not all the tuples in the normal skyline will be in the result, and so more the constraints are tight, smaller will be the output, here $\mathcal{F}$-Skyline has a benefit, since it's always possible to add more constraints to reduce the size, instead, in ε-skyline tuning the parameters could be difficult because there's could be a tradeoff between user preferences and output, and fine-tuning those parameters could be onerous.

Although even if the output size is not fixed, studies [1] showed the efficiency of flexible skyline, which can reduce the number of points in the skyline, in fact, it's easily possible to obtain an ND set with $< 10\%$ of tuples of the total skyline, and a PO with even a $< 1\%$ of the total tuples. Of course, increasing the number of elements in the dataset, the number of dimensions, and the constraint increase the efficacy of $\mathcal{F}$-skyline.

## 7.3 Generalization

Last, about generalization, since in ORD/ORU, as said, the output size is fixed, the amount of generalization will depend on the chosen number, and, since the optimal number of tuples cannot be known in advance, that's a drawback not easily resolvable, on the other hand, ε-skyline and flexible skyline, will have a better overview of the entire database.

$\mathcal{F}$-Skyline, with the fact that from a single dataset can retrieve 2 sets ND and PO, gives the user more options regarding how generalized the result must be, since it's possible to show both sets or just the PO which is more specific but still a lot smaller the entire non-dominated set of a normal skyline.

None of the above studied operators can reach the generalization obtained with the normal skyline operator, which remains the preferred one if a completely user agnostic analysis of the dataset is wanted; both $\mathcal{F}$-Skyline and ε-skyline can perform the same operations as Skyline if no constraint is applied and so they can replace the normal Skyline in any database implementation.

Graph 1- qualitative comparison

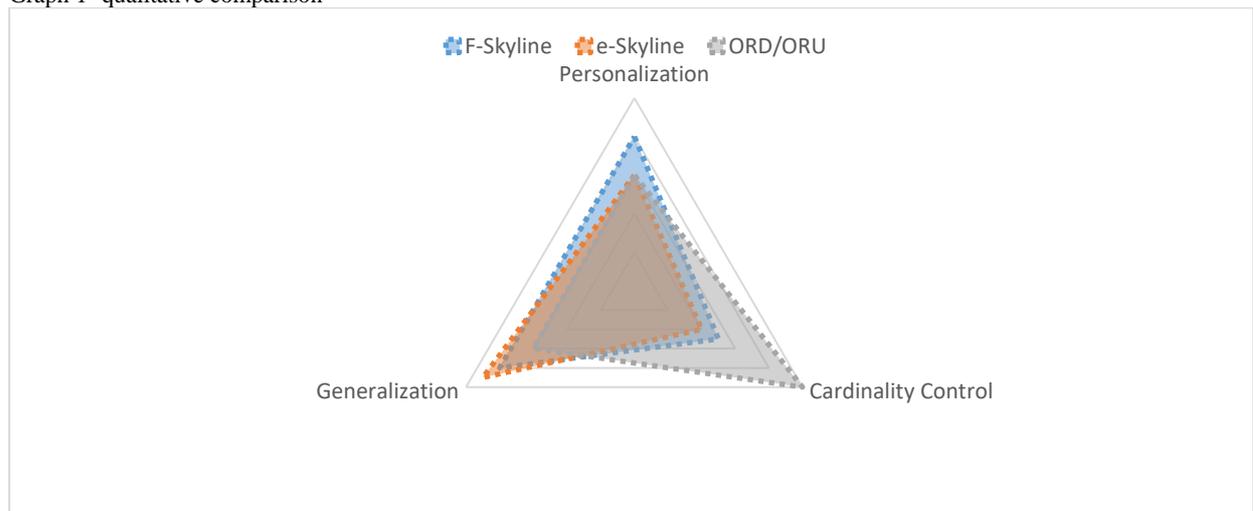

## 8 Conclusion

In this survey 3 of the most useful flexible/restricted skyline techniques have been analysed, a general view of the idea behind those has been given and the mainline to compute them, then a comparison between the shown techniques and how they are linked with the basic top-k and skyline operators, last to give a better general view a radar graph has been given which qualitative gives a comparison based on the difference exposed in chapter 7.

The Graph 1 assumes that Skyline has the maximum value in generalization but doesn't have any personalization or cardinality control, instead top-k is the opposite with maximum personalization and cardinality control, but no generalization, as shown in Table 1 – Features comparison

This graph can be used to choose which of the techniques is more suitable for the task that the user wants to fulfil.